\documentclass[a4paper]{article}
\usepackage{graphicx}
\usepackage{colordvi}
\usepackage{amsmath}
\usepackage{amssymb}
\usepackage{natbib}
\renewcommand{\vec}[1]{\mbox{\boldmath \(#1\)}}
\newcommand{\tensor}[1]{\mbox{\boldmath $#1$}}
\newcommand{\upd}{\mathrm{d}}
\newcommand{\tr}{\mathrm{tr}}
\makeatletter
\newcommand{\figcaption}[1]{\def\@captype{figure}\caption{#1}}
\newcommand{\tblcaption}[1]{\def\@captype{table}\caption{#1}}
\makeatother

\title{ Melt-Mixing by Novel Pitched-Tip Kneading Disks 
 \\
 in a Co-Rotating Twin-Screw Extruder
 }
\author{
Yasuya Nakayama\(^{1}\)\footnote{
\texttt{nakayama@chem-eng.kyushu-u.ac.jp}
}
,
Eiji Takeda\(^{1}\),
\\
Takashi Shigeishi\(^{2}\),
Hideki Tomiyama\(^{2}\),
and Toshihisa Kajiwara\(^{1}\)\footnote{
\texttt{kajiwara@chem-eng.kyushu-u.ac.jp}
}
\vspace*{1ex}
\\
{\normalsize\itshape\(^{1}\)Department of Chemical Engineering,
Kyushu University, Fukuoka 819-0395, Japan
}
\\
{\normalsize\itshape\(^{2}\)Hiroshima Plant, The Japan Steel Works Ltd.
1-6-1 Funakoshi-minami, Hiroshima 736-8602, Japan
}
}
\date{\empty}
\begin{document}
\maketitle
\begin{abstract}
Melt-mixing in twin-screw extruders is a key process in the development
of polymer composites. 
Quantifying the mixing performance of kneading elements based on their
internal physical processes is a challenging problem.
We discuss melt-mixing by novel kneading elements called ``pitched-tip
kneading disk~(ptKD)''.
The disk-stagger angle and tip angle are the main geometric parameters
of the ptKDs.
We investigated four typical arrangements of the ptKDs, which are
forward and backward disk-staggers combined with forward and backward
tips.
Numerical simulations under a certain feed rate and screw revolution
speed were performed, and the mixing process was investigated using
Lagrangian statistics.
It was found that the four types had different mixing characteristics,
and their mixing processes were explained by the coupling effect of drag
flow with the disk staggering and pitched-tip and pressure flows, which
are controlled by operational conditions.
 The use of a pitched-tip effectively to controls the balance of the
 pressurization and mixing ability.
\end{abstract}

\vspace*{0.5ex}
\vspace*{0.3ex}

\section{Introduction}
A twin-screw extruder~(TSE) is widely used in various industries,
including polymer processing, food processing, rubber compounding,
pharmaceutical development, and processing other highly viscous
materials because multiple processes, for instance, heating, melting,
compounding, blending, reaction, and devolatilization, can be 
conducted continuously in one operation.
In the plastic industry, a TSE is used to develop products with certain
desired properties by compounding and blending different types of
polymers, fillers, and other additives.
In this process, controlling the mixing in the TSE is critical to
optimize and improve the productivity and functionality of products.

The control of the mixing process is based on the design of the kneading
screw element.
In the past, among different kneading screws, kneading disks~(KDs) were
the most commonly used for general purpose compounding due to its mixing
efficiency.
One main geometric parameter of a KD is the disk-stagger angle that
controls both the pumping capability and inter-disk leakage flow
simultaneously. However, these two factors have different
effects on the mixing process.

To control these two factors separately, we have proposed a
novel design of a kneading screw element, called ``pitched-tip kneading
disk~(ptKD)''~\citep{shigeishi06:_devel_of_special_knead_screw}.
In the ptKDs, disk tips are pitched to the screw axes in contrast to the
tips that are pitched parallel to the screw axes in conventional KDs.
By arranging the pitched-tip angle along with the disk-stagger angle,
the pumping capability and the inter-disk leakage flow are controlled
independently. This flexibility makes it possible to better
control distributive mixing and dispersive mixing than with
conventional KDs.

For a specific application of novel screw or barrel designs, it is
difficult to optimize extrusion conditions because there are a large number
of parameters including geometric and operational parameters of the device
and rheological parameters of the materials, which affect the flow
pattern and mixing process.
It is difficult to experimentally obtain information on the flow pattern
and mixing process in a TSE because TSEs are operated under severe
conditions, including high temperature and high pressure; moreover, the
channel in TSEs in which materials are transported is highly
complicated.
Numerical simulation can aid in assessing material behaviors in a TSE.

Three-dimensional fluid dynamic simulations of melt-mixing in TSE have
been performed by several authors
\citep{kajiwara96:_numer_study_of_twin_screw,yoshinaga00:_mixin_mechan_of_three_tip,ishikawa00:_d_numer_simul_of_nonis,ishikawa00:_numer_simul_and_exper_verif,ishikawa01:_d_non_isoth_flow_field,Ishikawa2002Flow,funatsu02:_d_numer_analy_mixin_perfor,yang92:_flow_field_analy_of_knead,cheng98:_distr_mixin_in_convey_elemen,yao98:_influen_of_desig_disper_mixin,cheng97:_study_of_mixin_effic_in,bravo00:_numer_simul_of_press_and,bravo04:_study_of_partic_trajec_resid,jaffer00:_exper_valid_of_numer_simul,lawal95:_mechan_of_mixin_in_singl,lawal95:_simul_of_inten_of_segreg,kalyon07:_integ_approac_for_numer_analy,malik05:_finit_elemen_simul_of_proces,alsteens04:_param_study_of_mixin_effic}.
It is essential to perform three-dimensional simulations to observe the
material flow and the mixing process in TSE because the process is
mainly dominated by the three-dimensional structure of the channel.

In this article, we investigate the flow characteristics and mixing
process in melt-mixing with several types of ptKDs.
We conducted three-dimensional numerical simulations of melt-mixing zone
in a TSE.
Material kinetics was solved with the Lagrangian method, and the
characteristics of distributive and dispersive mixing in melt-mixing
with several types of ptKD was investigated.

\section{Pitched-tip kneading disk}
One element of a ptKD is composed of several kneading blocks, like the
conventional KD, but each kneading block of a ptKD has non-parallel tips
to the screw axes, in contrast to conventional KDs.
When a tip is pitched to pump fluid forward~(backward) along the screw
rotation, it is called a ``forward~(backward) tip''.
When the disk-stagger angle is arranged to pump fluid forward~(backward)
along the screw rotation, it is called a ``forward~(backward) stagger''.
According to these two main geometric characteristics, ptKDs are
typically classified as one of four types:
backward stagger and forward tip~(Bs-Ft),
forward stagger and forward tip~(Fs-Ft),
backward stagger and backward tip~(Bs-Bt), and
forward stagger and backward tip~(Fs-Bt).
Figure~\ref{fig:tkd_top_view} shows the top views of the four typical
types of ptKDs.

The four types of ptKDs in Fig.~\ref{fig:tkd_top_view} are composed of
five blocks of one element and are chosen for our investigation.
Based on the barrel diameter \(D=69.0\)~mm, each self-wiping
block~\citep{booy78:_geomet_of_fully_wiped_twin_screw_equip} had a
flight radius of 0.48\(D\) and a width of 0.3\(D\).
Disk-stagger angles were set to \(\pm 40.685^{\circ}\) for Fs-Ft and
Bs-Bt, \(\pm 49.315^{\circ}\) for Fs-Bt and Bs-Ft, respectively.  The
leads of four types of ptKDs were set to \(\pm 6D\).

\begin{minipage}[cbt]{\hsize}
\begin{tabular}[tb]{cc}
\begin{minipage}[cbt]{.5\hsize}
\center
\includegraphics[width=\hsize]{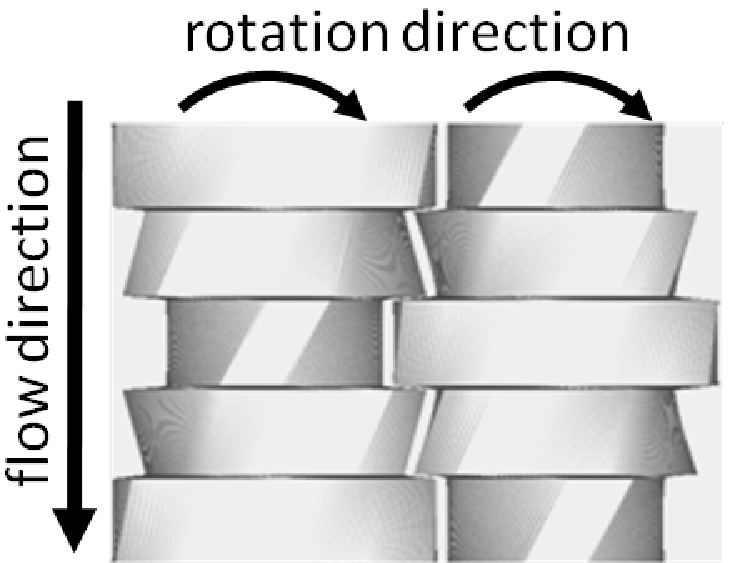} 
(a)
\end{minipage}
 & 
\begin{minipage}[cbt]{.5\hsize}
\center
\includegraphics[width=\hsize]{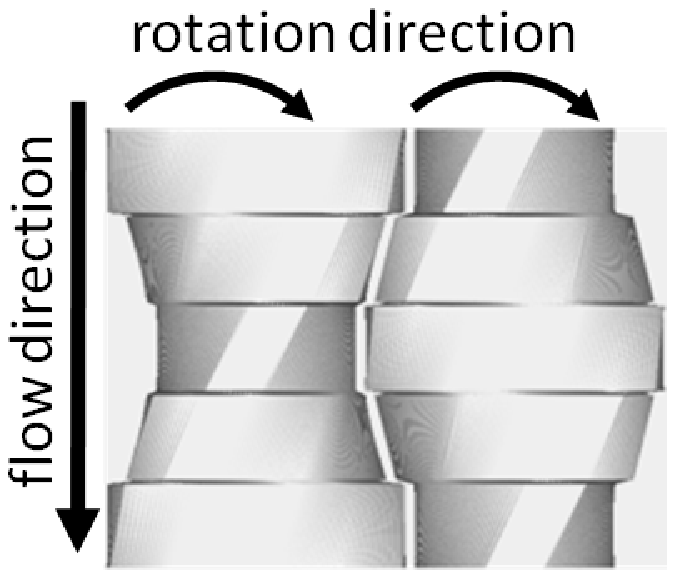} 
(b)
\end{minipage}
\\
\begin{minipage}[cbt]{.5\hsize}
\center
\includegraphics[width=\hsize]{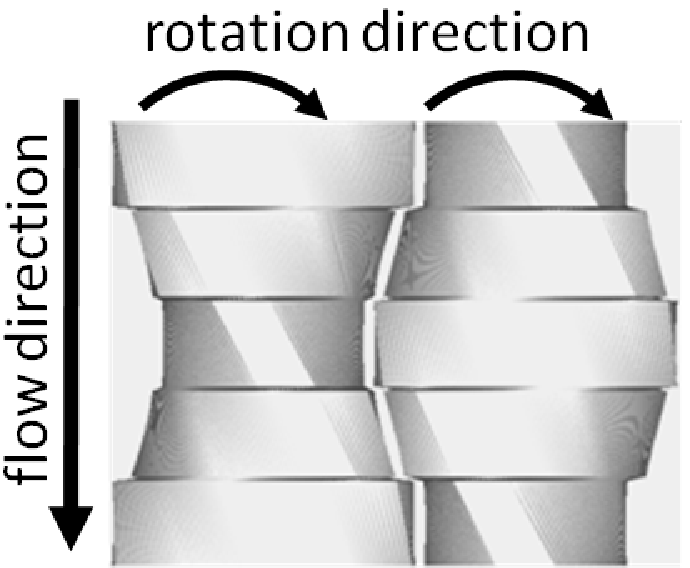} 
(c)
\end{minipage}
 & 
\begin{minipage}[cbt]{.5\hsize}
\center
\includegraphics[width=\hsize]{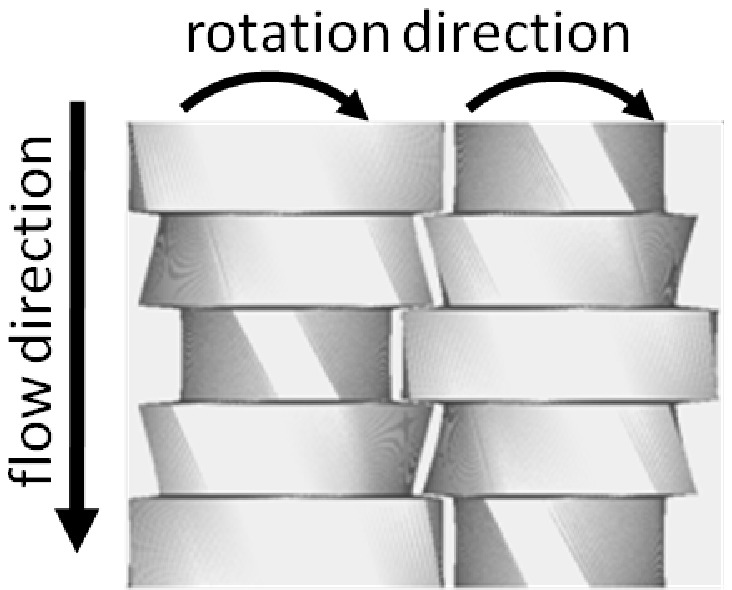} 
(d)
\end{minipage}
\end{tabular}
\figcaption{Top view of typical pitched-tip kneading disks: (a) backward
 stagger and forward tip~(Bs-Ft), (b) forward stagger and forward
 tip~(Fs-Ft), (c) backward stagger and backward tip~(Bs-Bt), (d) forward
 stagger and backward tip~(Fs-Bt).  }
\label{fig:tkd_top_view}
\end{minipage}
\section{Numerical simulation}
\subsection{Basic equations}
Our simulation focuses on the situation in which the material is fully
filled in the melt-mixing zone in a TSE.
The flow of the polymer melt in the TSE was assumed to be
incompressible, and the Reynolds number was assumed to be much less than
unity so that the inertial effects could be neglected.
We considered the pseudo-steady state of mass and heat transfers under
which the momentum and temperature fields instantly reach the steady
state for a given channel geometry, and the time evolution of the fluid
is totally governed by the screw rotation.
With these assumptions, the governing equations become
\begin{align}
 \vec{\nabla}\cdot\vec{v}&=0,
\label{eq:incompressibility}
\\
\vec{0} &= -\vec{\nabla}p +\vec{\nabla}\cdot\tensor{\tau},
\label{eq:stokes_equation}
\\
\rho C_{p}\vec{v}\cdot\vec{\nabla}T &= k \nabla^{2}T
 +\tensor{\tau}:\tensor{D},
\label{eq:temperature_equation}
\end{align}
where \(p\) and \(\tensor{\tau}\) are the hydrostatic pressure and the
deviatoric stress, respectively; \(\rho\), \(C_{p}\), \(T\) and \(k\)
are the mass density, the specific heat capacity, the temperature and
the thermal conductivity, respectively; and
\(\tensor{D}=\left[\vec{\nabla}\vec{v}+\left(\vec{\nabla}\vec{v}\right)^{T}\right]/2\)
is the strain-rate of the velocity field \(\vec{v}\), where \((.)^{T}\)
indicates the transpose.
\subsection{Working fluid and constitutive equations}
\begin{figure}[!htbp]
\includegraphics[width=\hsize]{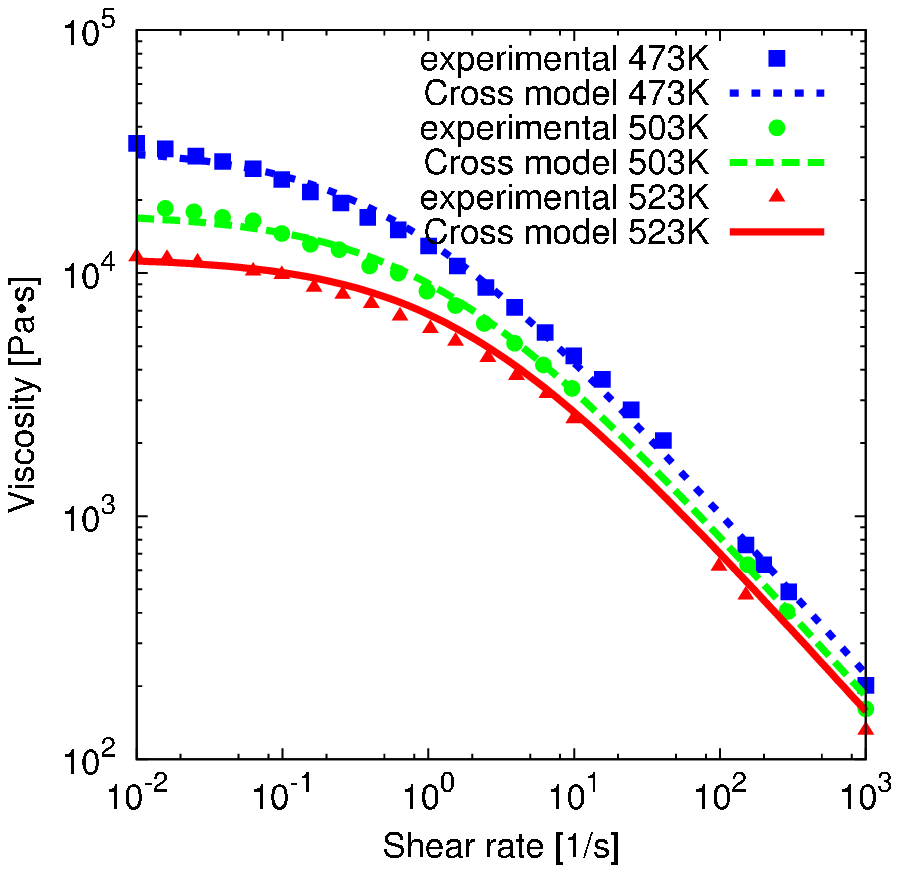}
\caption{Shear viscosity of a polypropylene
 melt\citep{ishikawa00:_d_numer_simul_of_nonis}. The solid lines are the
 Cross-exponential
 model~(\ref{eq:cross_model})-(\ref{eq:exponential_model}) fit.  The
 parameters are \(T_{0}=473\)~K, \(\lambda(T_{0})=1.70249\)~s,
 \(\eta_{0}(T_{0})=32783\)~Pa\(\cdot\)s, \(n(T_{0})=0.330793\)~(-),
 \(\beta=0.0208143\)~K\(^{-1}\).  }
\label{fig:cross_viscosity}
\end{figure}

We employ a homogeneous fluid as the working fluid because we are
focused on characterizing the melt-mixing of different screw geometries.
The working fluid is assumed to be a viscous shear-thinning fluid that
follows the Cross-exponential model:
\begin{align}
\tensor{\tau} &= 2\eta \tensor{D},
\label{eq:viscous_stress}
\\
%\begin{split}
 \eta(\dot{\gamma}, T)&=\frac{\eta_{0}(T_{0})H(T,T_{0})}{1+\left(
\lambda(T_{0}) H(T,T_{0})\dot{
\gamma}
\right)^{1-n}},
\label{eq:cross_model}
\\
H(T,T_{0}) &=\exp\left[-\beta(T-T_{0})\right],
\label{eq:exponential_model}
\\
\dot{\gamma} &= \sqrt{2\tensor{D}:\tensor{D}},
%\end{split}
%\label{eq:cross_exponential}
\end{align}
The parameters of the Cross-exponential model were determined by fitting
function to the experimental shear viscosity of a polypropylene
melt~\citep{ishikawa00:_d_numer_simul_of_nonis}, as shown in
Fig.\ref{fig:cross_viscosity}.

The mass density, specific heat capacity, and thermal conductivity were
assumed to be temperature-independent in the simulation:
\(\rho=735.0\)~kg/m\(^{3}\), \(C_{p}=2100.0\)~J/(kg\(\cdot\)K), and
\(k=0.15\)~W/mK, respectively.

\subsubsection{Boundary conditions}	
The no-slip condition on the velocity at the barrel and screw surfaces
is assumed.
The velocities at inlet and outlet boundaries were obtained according to
the same cross-sectional velocity profile under the given volumetric
flow rate of \(Q=60\)~cm\(^3\)/s~(\(\approx\)159~kg/h).
The temperatures on the barrel surface and at the inlet boundary were
set to 473~K and 453~K, respectively. The natural boundary conditions
for the temperature equation in the exit boundary plane and the screw
surface were assigned.

\subsubsection{Penalty method for the incompressibility condition}
To incorporate the incompressibility condition
(\ref{eq:incompressibility}) in momentum conservation
(\ref{eq:stokes_equation}), the penalty method is utilized: the pressure
in Eq.(\ref{eq:stokes_equation}) is eliminated by imposing the relation
\begin{align}
 p&=-\lambda \eta_{0}\vec{\nabla}\cdot\vec{v},
\label{eq:penalty_pressure}
\end{align}
with a constant \(\lambda\gg
1\)\citep{zienkiewicz77:_finit_elemen_method}.  The penalty term in the
momentum conservation effectively acts as a very large bulk viscous
stress; therefore, the resulting velocity field satisfies the
incompressibility condition (\ref{eq:incompressibility}) in the limit
as \(\lambda \to\infty\).

\subsubsection{Spatial discretization and time evolution}
The channel in a TSE is decomposed by the hexahedral element.  The velocity
and the temperature are approximated by tri-quadratic interpolation with
the 27-node in each hexahedral element.
The momentum equations (\ref{eq:stokes_equation}) and
(\ref{eq:penalty_pressure}) were discretized by the Galerkin finite
element method.
The temperature equation (\ref{eq:temperature_equation}) was discretized
by streamline-upwind/Petrov--Galerkin
method~\citep{brooks82:_stream_upwin_petrov_galer_formul,marchal87:_new_mixed_finit_elemen_for}
to stabilize the advection term.

The time evolution of the velocity and temperature fields was constructed
with the converged fields for every three degrees of screw rotation.
This computation was the most costly.  For reference, one simulation
takes about 68 hours on a computer with an
Intel\(^{\text{\textregistered}}\) Core\(^{\text{TM}}\) 2 Extreme
processor.

\subsubsection{Lagrangian point tracer}
The trajectories of the passive tracers were determined based on the
solved velocity field.  The time evolution of the tracer density was
utilized to assess the flow pattern in the melt-mixing zone with a
ptKD. The set of tracer trajectories was utilized to compute the residence
time distribution~\citep{levenspiel98:_chemic_react_engin_edition} and
to investigate the flow history and mixing process.
The Lagrangian-history average of a quantity \(f\) over the trajectory of
\(\alpha\)th tracer is defined as
\begin{align}
 \overline{f_{\alpha}}^{T_{\alpha}} &=\frac{1}{T_{\alpha}}\int_{0}^{T_{\alpha}}\upd
 s\int\upd\vec{x}\delta\left(
\vec{x}-\vec{X}_{\alpha}(s)
\right)f(\vec{x},s),
\label{eq:lagrangian_average_of_f}
\end{align}
where \(T_{\alpha}\) and \(\vec{X}_{\alpha}(.)\) are the residence time
and position of the \(\alpha\)th tracer, respectively, and \(\delta(.)\)
is the Dirac delta distribution.
The statistical distribution of \(\overline{f_{\alpha}}^{T_{\alpha}}\)
characterizes the process in the melt-mixing zone.

Initially, 2000 points were uniformly distributed in a certain section
relative to the axial position, which is arbitrarily set in the second disk.
They were advected until they reach the outlet section.
When computing the tracer advection, some tracers went out of bounds
because the time resolution of the velocity field was limited.  To
circumvent the effect of the lost tracers on the statistics, we set the
number of initial points of 2000, which ensured that a sufficient number
of points reached the outlet. In our advection computation, at least 93\% of
the tracers reached the outlet.
\section{Results and discussion}
\subsection{Basic characteristics of ptKDs}
\subsubsection{Drag ability of ptKD}
\begin{figure}[!htbp]
\center
\includegraphics[width=\hsize]{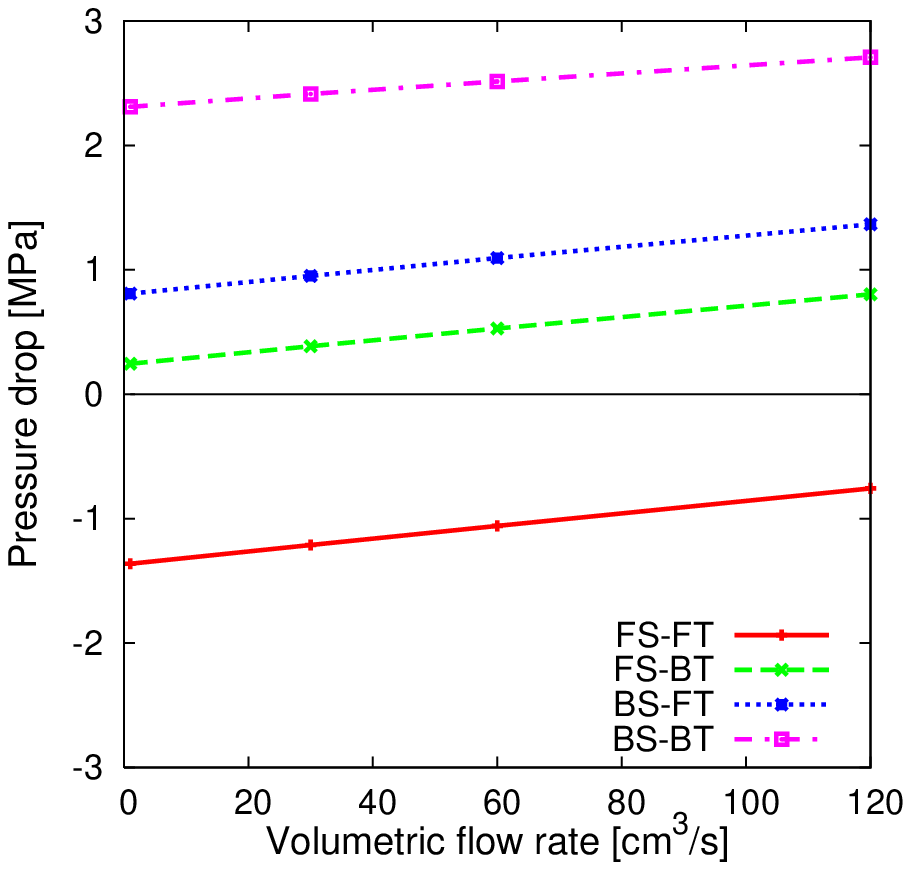} 
\caption{Pressure drop averaged over screw rotation with a screw
 rotation speed of 200~rpm as a function of the volumetric flow rate.}
\label{fig:pressure_drop}
\end{figure}
To assess the effect of pitched tips on the pressurization capability,
the pressure drop from the inlet to the outlet was calculated under a
fixed screw rotation speeds and various volumetric flow
rates~(Fig.\ref{fig:pressure_drop}).
For a fixed volumetric flow rate, the pressure drop was largest in
Bs-Bt, followed by Bs-Ft, Fs-Bt, and Fs-Ft.
The results in Fig.\ref{fig:pressure_drop} show that the disk
stagger mainly controls the drag ability of ptKD and that the
pitched tips only experience slight modification.
It is noted that the pressure drop in Fs-Ft ptKD was negative for a
certain range of the volumetric flow rate, which indicates that the
pressure-driven flow on average occurred in the direction from the
outlet to the inlet.

\subsubsection{Temperature profile}
\begin{figure}[!htbp]
\center
\includegraphics[width=\hsize]{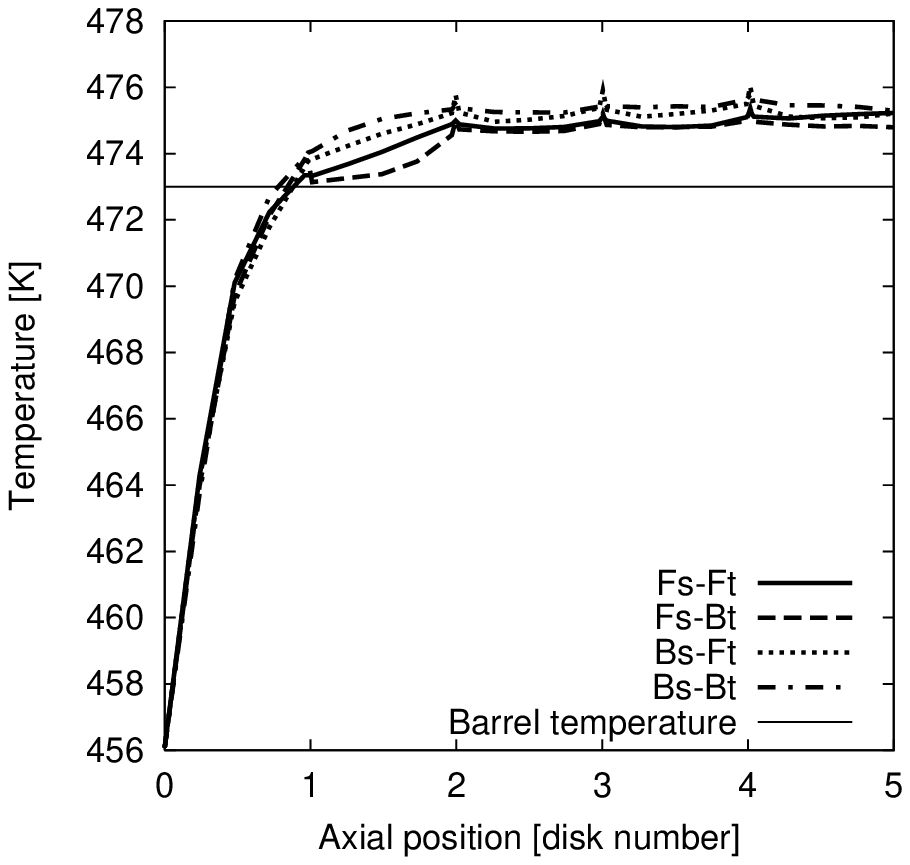} 
\caption{Average temperature profile along the axial direction 
with a volumetric flow rate of
 60~cm\(^{3}\)/s and a screw rotation speed of 200~rpm.}
\label{fig:section_time_averaged_temperature}
\end{figure}
Figure~\ref{fig:section_time_averaged_temperature} shows the average
temperature profile along the extrusion direction.  For all types of
ptKDs, the temperature rose from the inlet temperature to the barrel
temperature over one block. Next, the average melt temperature increased
two or three degrees beyond the barrel temperature due to viscous
heating effects.  Because the difference in the results with different
types of ptKD was very small, we assume that the temperature did not
have a significant effect on the mixing ability.

\subsubsection{Residence time and Lagrangian history of exerted stress}
\label{sec:stress_exit_time}
\begin{figure}[!htbp]
\includegraphics[width=\hsize]{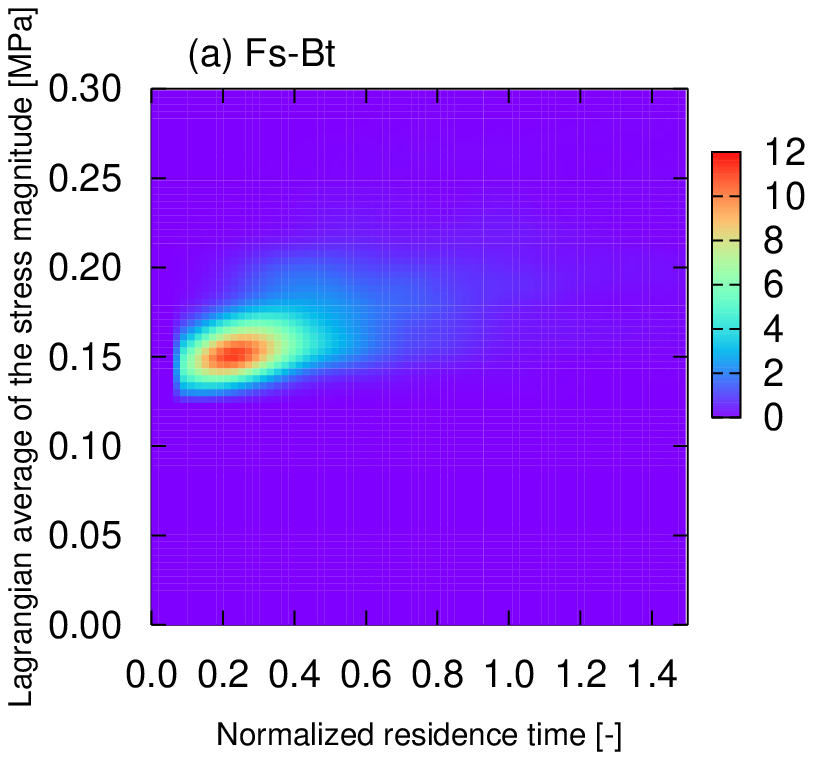} 
\includegraphics[width=\hsize]{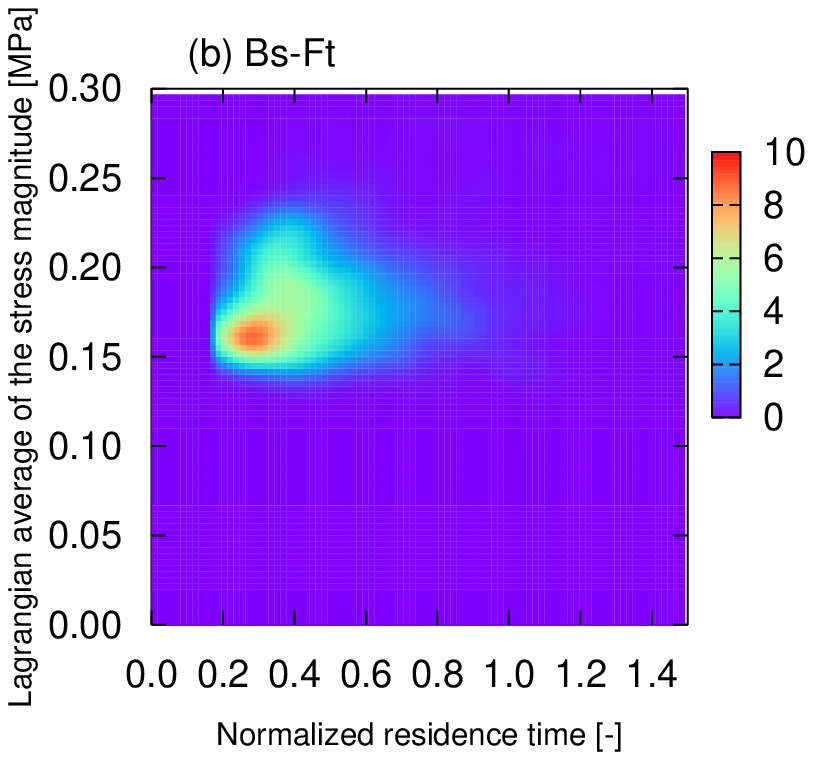} 
\caption{Joint probability density of the residence time and
 Lagrangian-history average of the stress magnitude of
 Eq.(\ref{eq:mises_stress}) with a volumetric flow rate of
 60~cm\(^{3}\)/s and a screw rotation speed of 200~rpm:
(a) Fs-Bt ptKD, and (b) Bs-Ft ptKD.
}
\label{fig:jointpdf_residence_time_average_stress_fs_bt_bs_ft}
\end{figure}

\begin{figure}[!htbp]
\includegraphics[width=\hsize]{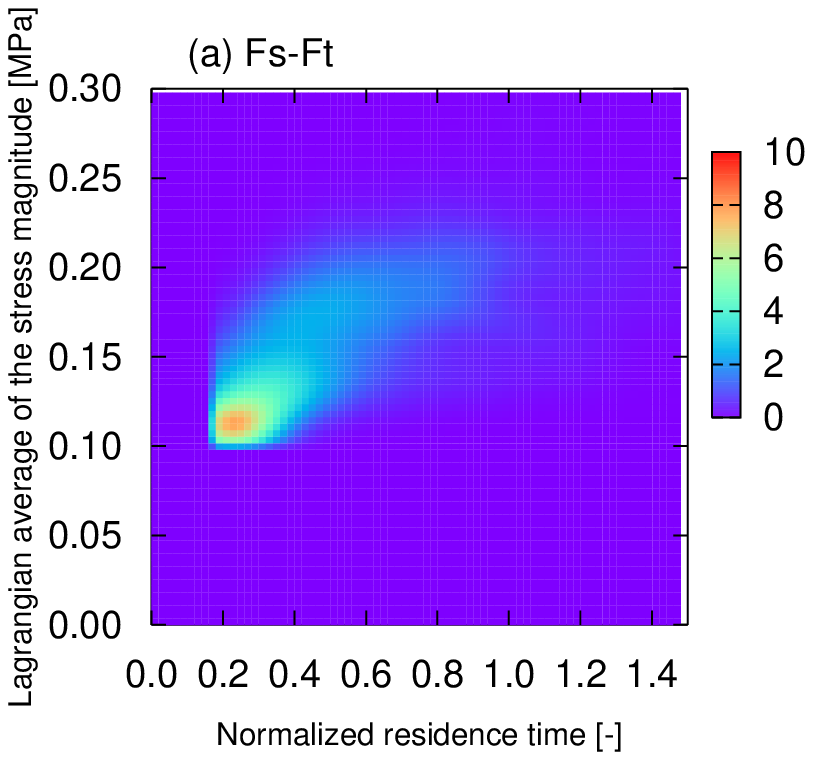} 
\includegraphics[width=\hsize]{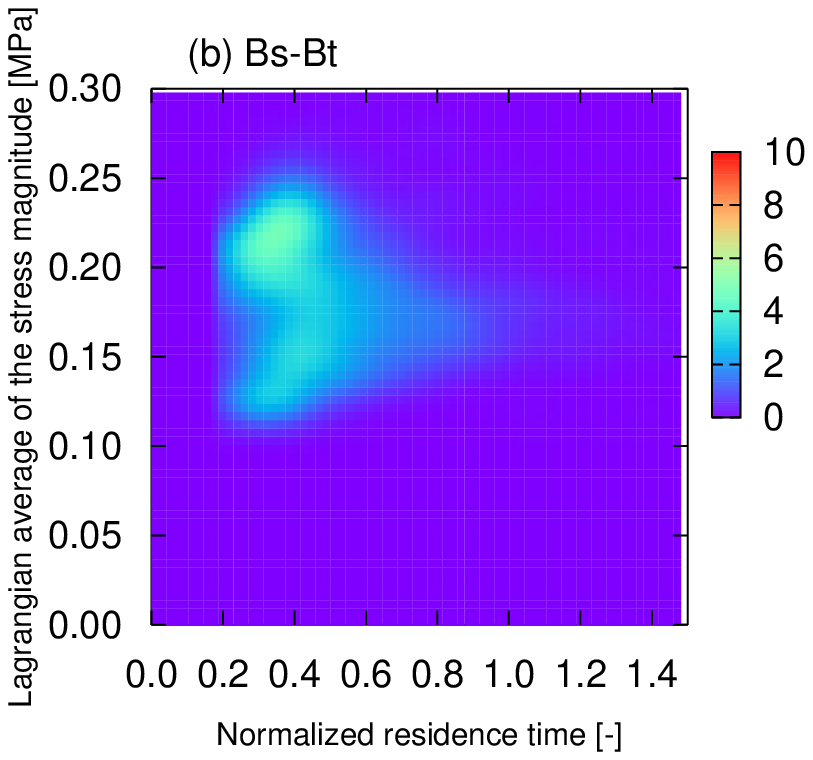} 
\caption{ The same plot as
Fig.\ref{fig:jointpdf_residence_time_average_stress_fs_bt_bs_ft} but for
(a) Fs-Ft ptKD and (b) Bs-Bt ptKD.}
\label{fig:jointpdf_residence_time_average_stress_fs_ft_bs_bt}
\end{figure}

Figures~\ref{fig:jointpdf_residence_time_average_stress_fs_bt_bs_ft} and
\ref{fig:jointpdf_residence_time_average_stress_fs_ft_bs_bt} show the
joint probability density function~(PDF) of the residence time and the
Lagrangian-history average of stress magnitude under
\(Q=60\)~cm\(^{3}\)/s and 200~rpm.
To measure the stress magnitude, a second-order invariant of
\(\tensor{\tau}\) is considered
\footnote{
Note that 
although
one of the authors (T.K.)~
defined 
``characteristics shear
stress'' as
\begin{align*}
 \sqrt{4\left(\tau_{xy}^{2}+\tau_{yz}^{2}+\tau_{zx}^{2}\right)
+\left(\tau_{xx}-\tau_{yy}\right)^{2}
+\left(\tau_{yy}-\tau_{zz}\right)^{2}
+\left(\tau_{zz}-\tau_{xx}\right)^{2}
},
\end{align*}
~\citep{kajiwara96:_numer_study_of_twin_screw} which was used in
\citet{yoshinaga00:_mixin_mechan_of_three_tip} and
\citet{ishikawa01:_d_non_isoth_flow_field,Ishikawa2002Flow},
it is not
an invariant of a second-rank tensor; therefore, it is not adequate to quantify the
magnitude of the stress tensor.}.
Although \(\tensor{\tau}=2\eta\tensor{D}\) is traceless by the
incompressibility condition~(\ref{eq:incompressibility}), a small but
finite \(\tr\tensor{\tau}\) is unavoidable in numerical simulations with
finite arithmetic. Therefore, we use the traceless part of
\(\tensor{\tau}\) to compute the stress magnitude,
\(\sigma\), as
\begin{align}
 \sigma &=\sqrt{\frac{3}{2}\tensor{\tau}':\tensor{\tau}'},
\label{eq:mises_stress}
\\
 \tensor{\tau}' &\equiv \tensor{\tau}-\frac{1}{3}(\tr\tensor{\tau})\tensor{I},
\end{align}
where \(\tensor{I}\) is the unit tensor.

\paragraph{Counter-pump combinations: Fs-Bt and Bs-Ft}
In Fs-Bt and Bs-Ft, the pitched-tip angle and disk-stagger angle are in
opposite directions, which we call the counter-pump combination.
In the counter-pump combination, the joint PDFs of \(T_{\alpha}\) and
\(\overline{\sigma_{\alpha}}^{T_{\alpha}}\) are almost
unimodal~(Fig.\ref{fig:jointpdf_residence_time_average_stress_fs_bt_bs_ft}).
These ptKDs are characterized by a broad residence time distribution and
a rather small fluctuation of
\(\overline{\sigma_{\alpha}}^{T_{\alpha}}\).
A small fluctuation of \(\overline{\sigma_{\alpha}}^{T_{\alpha}}\)
indicates that the average stress history of each trajectory is almost
homogeneous; thus, homogeneous dispersive mixing can be expected for
Fs-Bt and Bs-Ft.

\paragraph{Co-pump combinations: Fs-Ft and Bs-Bt}
In Fs-Ft and Bs-Bt, the pitched-tip angle and disk-stagger angle are in
the same direction, which we call the co-pump combination.
In the co-pump combination, the joint PDFs of \(T_{\alpha}\) and
\(\overline{\sigma_{\alpha}}^{T_{\alpha}}\) are spread over the plane
(Fig.\ref{fig:jointpdf_residence_time_average_stress_fs_ft_bs_bt}),
which indicates that the residence time and stress history of each
tracer are highly inhomogeneous.

For the
Fs-Ft~(Fig.\ref{fig:jointpdf_residence_time_average_stress_fs_ft_bs_bt}(a)),
the main peak indicates that large population of the tracer has a
shorter residence time and smaller average stress, which indicate that
pipeline flow occurs where tracers extrude under fast velocity and a
small strain-rate and mixing with surrounding fluid rarely occur.
The other trajectories have larger average stress during longer
residence time.

For the
Bs-Bt~(Fig.\ref{fig:jointpdf_residence_time_average_stress_fs_ft_bs_bt}(b)),
the joint PDF of \(T_{\alpha}\) and
\(\overline{\sigma_{\alpha}}^{T_{\alpha}}\) is bimodal;
although, the residence time fluctuation is smaller than those of other ptKDs.
The Lagrangian-history average of the stress magnitude has much
greater fluctuation than other ptKDs.
The bimodal structure indicates that pipeline flow also occurs in Bs-Bt,
but the group velocities of the pipeline and its surroundings are similar.
Note that the indication of the pipeline flow cannot be observed based
only on the residence time distribution. The joint PDF reveals the
highly inhomogeneous tracer history in the co-pump combination.

\subsection{Transport mechanism and characterization of mixing
  capability}
\subsubsection{Time evolution of tracer density}
\label{sec:axial_transport}
Spatial density of the tracer as a function of time, \(p(\vec{x},t)\),
is computed from the trajectories of the tracers,
\(\vec{X}_{i}(t)~(i=1,\ldots,N)\). To investigate the axial tracer
kinetics, the tracer density is projected in the axial~(\(z\)-)
direction as
\begin{align}
 p(\vec{x},t)&=\lim_{N\to \infty}\frac{1}{N}\sum_{i=1}^{N}\delta\left(
\vec{x}-\vec{X}_{i}(t)
\right),
\\
p(z, t) &= \int\upd x \upd y p(\vec{x},t).
\end{align}
Figure~\ref{sec:axial_transport} shows the evolution of the axial
distribution for the four ptKDs, provides detailed information on the
transport mechanism, and explains the stress-residence-time joint PDF
in the previous section.%~\ref{sec:stress_exit_time}.

RTD has been discussed often in connection with the axial
mixing~\citep{Ishikawa2002Flow,zhang09:_numer_simul_and_exper_valid}. However,
a detailed investigation of the axial transport for ptKDs revealed that
the source of the residence time fluctuation differs among the four
types of ptKDs, and that broadness of RTD does not necessarily means
good mixing capability.
\\
\begin{minipage}[cbt]{\hsize}
\begin{tabular}[tb]{cc}
\begin{minipage}[cbt]{.5\hsize}
\includegraphics[width=\hsize]{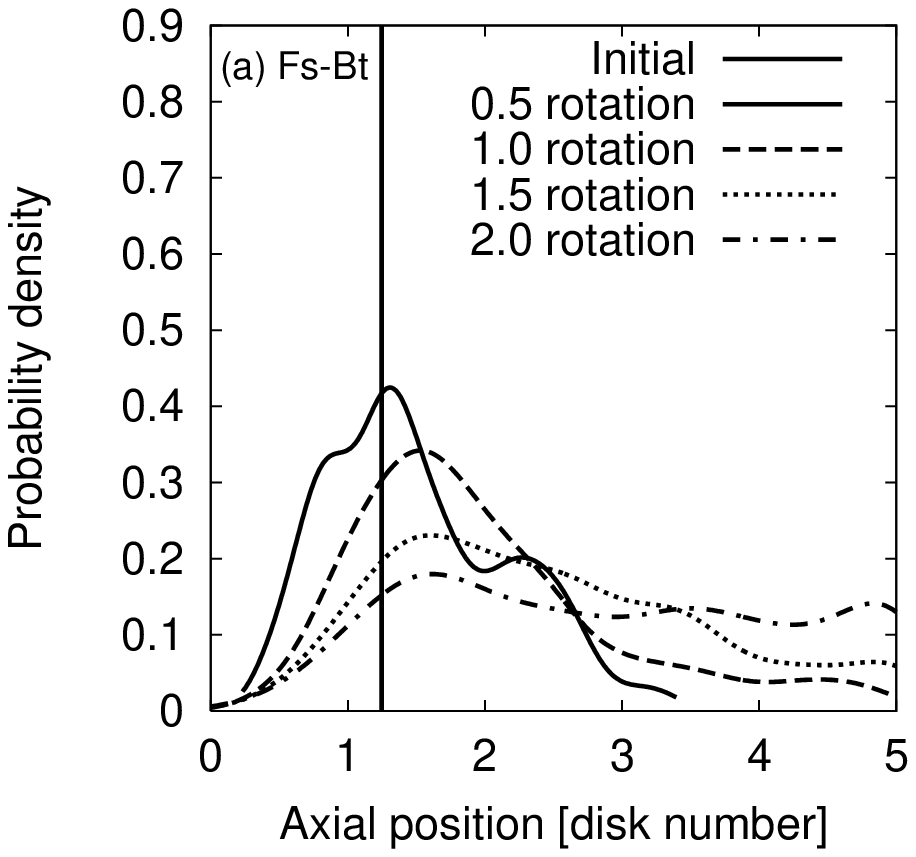} 
\end{minipage}
 & 
\begin{minipage}[cbt]{.5\hsize}
\includegraphics[width=\hsize]{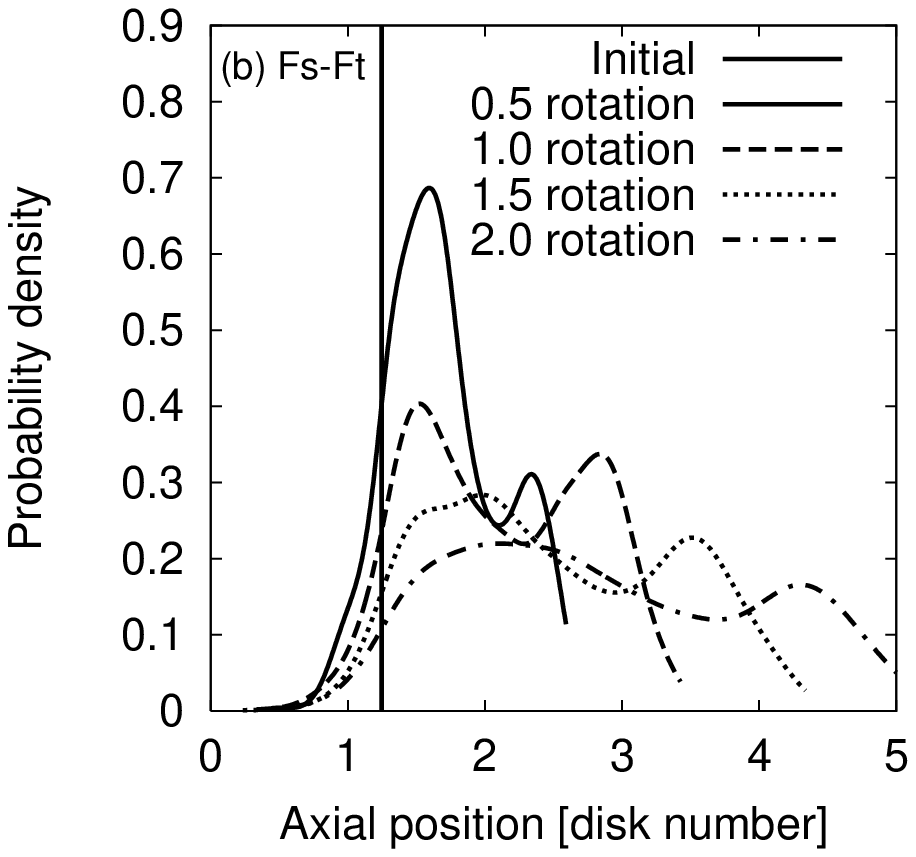} 
\end{minipage}
\\
\begin{minipage}[cbt]{.5\hsize}
\includegraphics[width=\hsize]{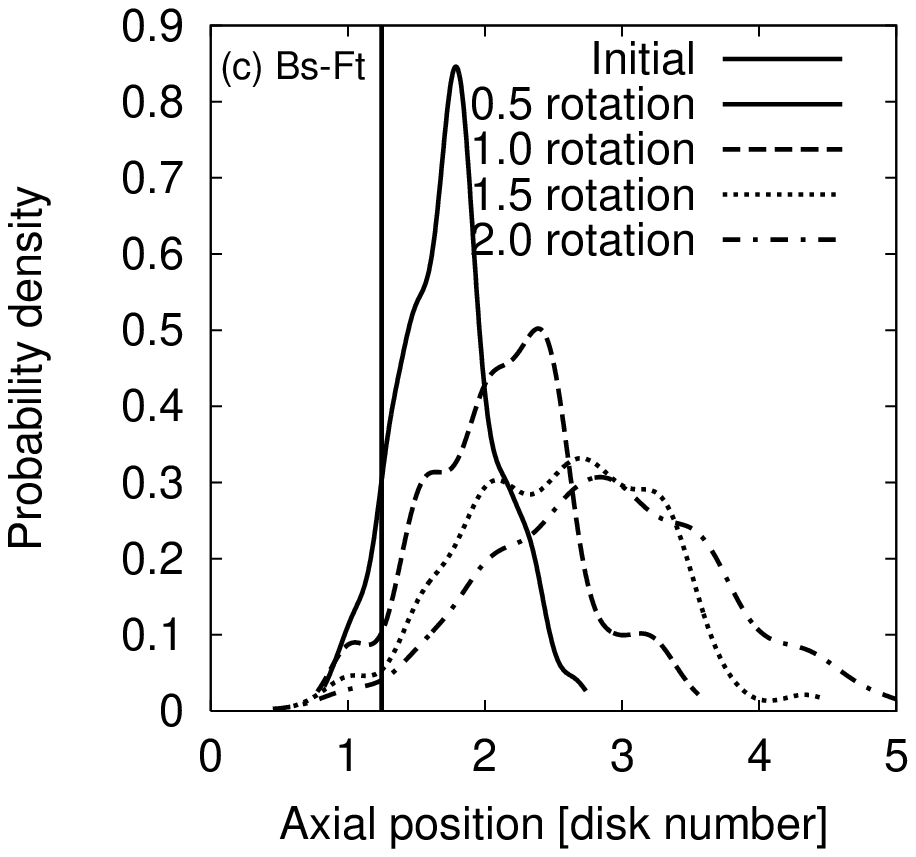} 
\end{minipage}
 & 
\begin{minipage}[cbt]{.5\hsize}
\includegraphics[width=\hsize]{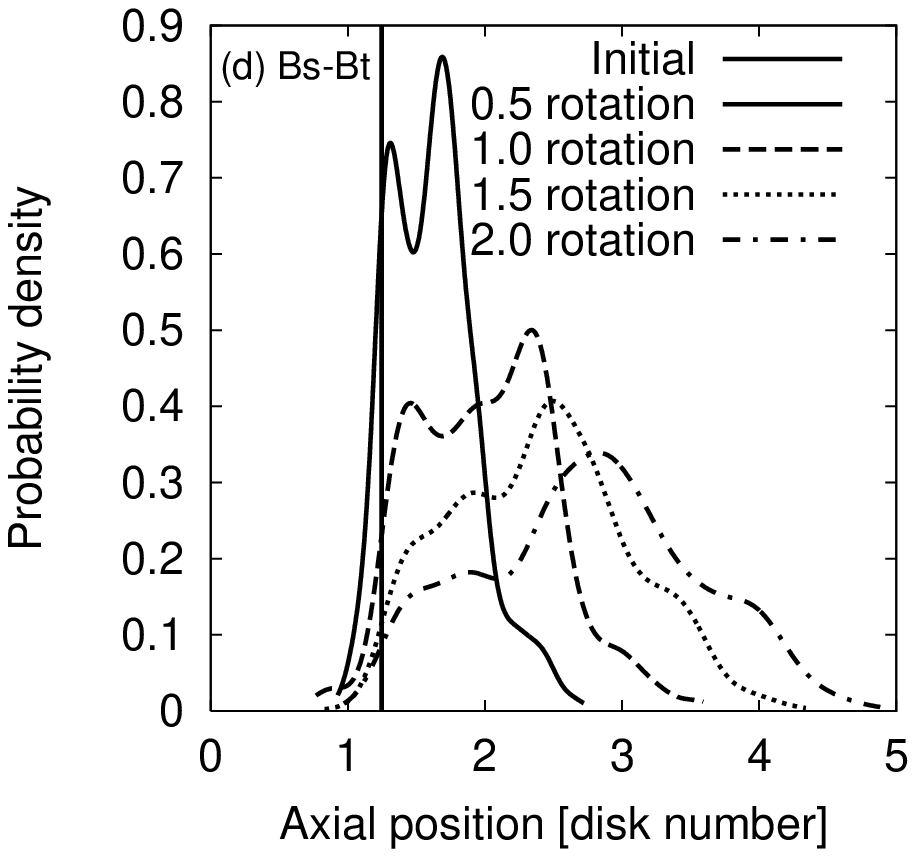} 
\end{minipage}
\end{tabular}
\figcaption{Time evolution of the tracer density on the axial direction for
(a) Fs-Bt,
(b) Fs-Ft, 
(c) Bs-Ft, and
(d) Bs-Bt.
The initial axial position of the tracer is depicted by the vertical line on
the second disk.
The different times correspond to the different lines indicated in the legend.
}
\label{fig:pdf_axial}
\end{minipage}
\paragraph{Counter-pump combinations: Fs-Bt and Bs-Ft}
For the counter-pump combination, the inter-disk leakage flow in front
of the pitched-tips enhanced the axial mixing.
As for Fs-Bt in Fig.\ref{fig:pdf_axial}(a), a portion of the tracer was
pumped backward at the inter-disk gap. 
A close investigation of each tracer's trajectory revealed that the
compression flow in front of the backward tips caused this inter-disk
backward pump and enhanced the axial mixing.
The effect of backward pitched-tips and circumferential mixing at the
intermeshing region promoted gross mixing over a whole channel of
melt-mixing zone.

As for Bs-Ft in Fig.\ref{fig:pdf_axial}(c), a portion of the tracer was
pumped forward at the inter-disk gap by the forward pitched-tips.  This
effect also caused axial mixing as in the case of Fs-Bt.
The broad RTDs observed in
Fig.\ref{fig:jointpdf_residence_time_average_stress_fs_bt_bs_ft} are the
result of axial mixing, and the unimodal stress-history is a result of
gross mixing.

\paragraph{Co-pump combinations: Fs-Ft and Bs-Bt}
For Fs-Ft in Fig.\ref{fig:pdf_axial}(b), 
it is observed that the unimodal axial density immediately split
into a bimodal density.
These two groups evolved with different axial group velocities.
A close investigation of each tracer's trajectory revealed that
the faster axial transport occurred at the roots of the kneading disks where 
it existed larger space to enhance the pressure-driven flow.
The faster group had a pipeline flow characteristic in which mixing with
surrounding fluid rarely occurred.
Because the forward-pump ability of Fs-Ft ptKD is large, a negative
pressure drop is generated in the volumetric flow
rate~(Fig.\ref{fig:pressure_drop}). The gross axial transport is a
result of the competing drag and pressure gradient.
The broader RTD in
Fs-Ft~(Fig.\ref{fig:jointpdf_residence_time_average_stress_fs_ft_bs_bt}(a))
is a result of pipeline flow and the surroundings.
Because the history of exerted stress was different between the pipeline
region and the surroundings, \(\overline{\sigma_{\alpha}}^{T_{\alpha}}\)
fluctuated significantly, as observed in
Fig.\ref{fig:jointpdf_residence_time_average_stress_fs_ft_bs_bt}(a).

As for Bs-Bt in Fig.\ref{fig:pdf_axial}(d), 
splitting of the axial density was also observed as in Fs-Ft, but the
group axial velocities were closer.
A close investigation of each tracer's trajectory revealed that
one group had a character of pipeline flow as in Fs-Ft.
The large fluctuation of \(\overline{\sigma_{\alpha}}^{T_{\alpha}}\)~(
Fig.\ref{fig:jointpdf_residence_time_average_stress_fs_ft_bs_bt}(b)) was
a result of pipeline flow and the surroundings.
However, similar axial group velocities for the pipeline and its
surroundings resulted in a narrower
RTD~(Fig.\ref{fig:jointpdf_residence_time_average_stress_fs_ft_bs_bt}(b)).

\subsubsection{Mixing capability: strain-rate type and maximum
   line-stretching rate}
For effective mixing, two points that are close initially must separate
rapidly over time to promote points distribution in space, stretch a
line element, and stretch an area element.
To quantify the mixing capability, 
we considered two aspects of this kinetic process:
the potential line-stretching capability and the type of
strain-rate field in which the line-stretching occurs.

\begin{figure}[!htbp]
\includegraphics[width=\hsize]{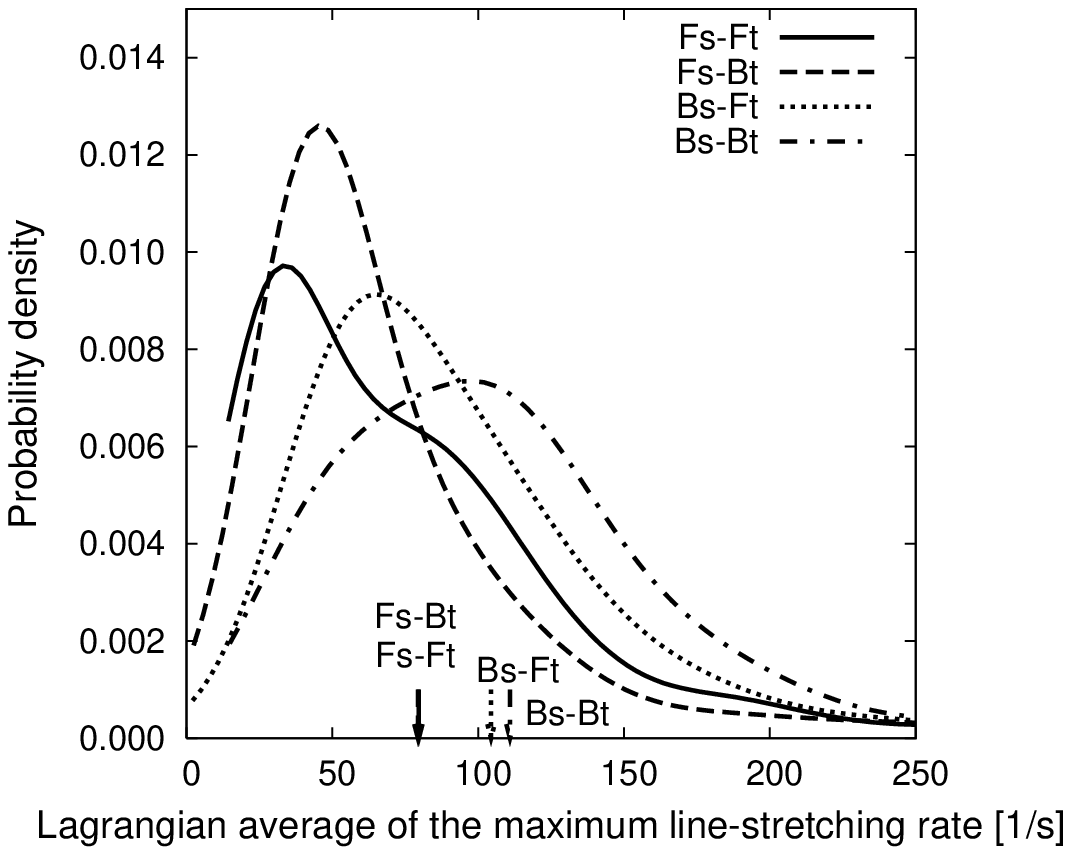} 
\caption{Probability density of the Lagrangian-history average of the
 maximum line-stretching rate with a the volumetric flow rate of
 60~cm\(^{3}\)/s and a screw rotation speed of 200~rpm. Arrows on the
 abscissa indicate the mean values.}
\label{fig:average_stretching_rate}
\end{figure}
To quantify the potential line-stretching capability, we used the
maximum line-stretching rate
\begin{align}
 \lambda_{M}(\vec{x},t)&=\max_{\vec{n},\left|\vec{n}\right|=1}\tensor{D}:\vec{n}\vec{n}.
\label{eq:maximum_stretching_rate}
\end{align}
The details of the quantity are
described in \ref{sec:def_flow_type_stretching_rate}.
The statistical distribution of the Lagrangian-history average of the
maximum line-stretching rate,
\(\overline{\left(\lambda_{M}\right)_{\alpha}}^{T_{\alpha}}\),
is depicted in Fig.\ref{fig:average_stretching_rate}
For co-pump ptKDs, namely Fs-Ft and Bs-Bt, PDFs of
\(\overline{\left(\lambda_{M}\right)_{\alpha}}^{T_{\alpha}}\) are
broad. In particular, for Fs-Ft, a shoulder at large
\(\overline{\left(\lambda_{M}\right)_{\alpha}}^{T_{\alpha}}\) is
observed.
The large fluctuation of
\(\overline{\left(\lambda_{M}\right)_{\alpha}}^{T_{\alpha}}\) is a
result of the pipeline flow.
In the trajectory by the pipeline flow,
\(\overline{\left(\lambda_{M}\right)_{\alpha}}^{T_{\alpha}}\) is small,
and the larger
\(\overline{\left(\lambda_{M}\right)_{\alpha}}^{T_{\alpha}}\) comes from
the surroundings.
The large fluctuation of 
\(\overline{\left(\lambda_{M}\right)_{\alpha}}^{T_{\alpha}}\) indicates
that gross mixing might be inhomogeneous in co-pump ptKDs.
It is noted that the fingerprint of the pipeline flow is clearer in the
PDF of the maximum line-stretching rate than in that of the residence time.

For counter-pump ptKDs, namely Fs-Bt and Bs-Ft, PDFs of
\(\overline{\left(\lambda_{M}\right)_{\alpha}}^{T_{\alpha}}\) are
unimodal, which indicates that no special structure exists in the flow
pattern in counter-pump ptKDs, and mixing over whole channel is expected.

It is noted that the mean values of
\(\overline{\left(\lambda_{M}\right)_{\alpha}}^{T_{\alpha}}\) indicated
in Fig.\ref{fig:average_stretching_rate} are determined by the
disk-stagger direction and are useless for characterizing the mixing
capability. To characterize the mixing capability, the fluctuation of
\(\overline{\left(\lambda_{M}\right)_{\alpha}}^{T_{\alpha}}\) should be
taken into account.

\begin{figure}[!htbp]
\includegraphics[width=\hsize]{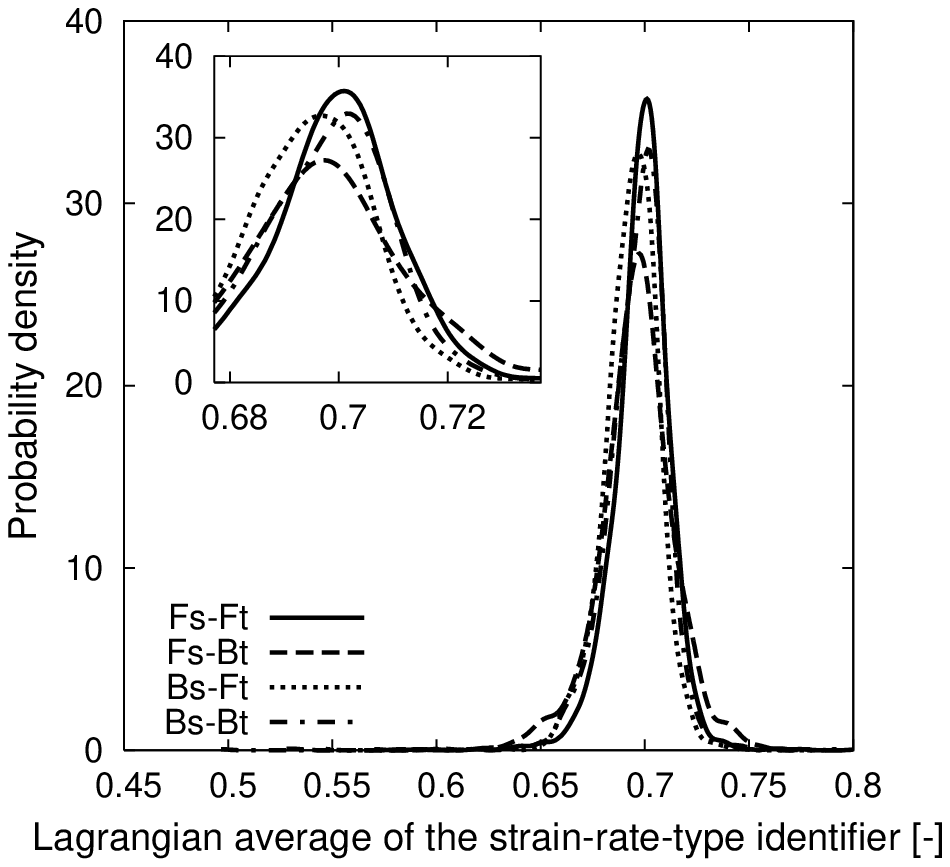} 
\caption{Probability density of the Lagrangian-history average of the
 strain-rate-type identifier with a volumetric flow rate of
 60~cm\(^{3}\)/s and a screw rotation speed of 200~rpm.}
\label{fig:average_mixing_efficiency}
\end{figure}
To identify a type of strain-rate field, the strain-rate-type identifier
as a function of the invariants of the strain rate,
\(\tensor{D}(\vec{x},t)\), is introduced as
\begin{align}
 f_{D}(\vec{x},t)&=\frac{\max_{\vec{n}}\tensor{D}:\vec{n}\vec{n}}{\sqrt{\tensor{D}:\tensor{D}}},
\label{eq:def_flow_type_indicator}
\end{align}
where \(\vec{n}\) is a unit vector.  The strain-rate-type identifier takes
values specific to the strain-rate state in incompressible flow,
\begin{align}
 f_{D}&=
 \begin{cases}
  \sqrt{2/3}& \text{(uniaxial elongational flow)}
\\
  \sqrt{1/2}& \text{(planar shear flow)}
\\
  \sqrt{1/6}& \text{(biaxial elongational flow)}
 \end{cases}
\label{eq:flow_type_indicator}
\end{align}
The details of the strain-rate-type identifier are summarized in
\ref{sec:def_flow_type_stretching_rate}.

The statistical distribution of the Lagrangian-history average of the
strain-rate-type identifier,
\(\overline{\left(f_{D}\right)_{\alpha}}^{T_{\alpha}}\), is depicted in
Fig.\ref{eq:def_flow_type_indicator}.
The modes of the PDFs for the four ptKDs are located near
\(\overline{\left(f_{D}\right)_{\alpha}}^{T_{\alpha}}=\sqrt{1/2}\approx
0.707\), 
which is a general characteristic of screw extruders in which the
contribution from 
circumferential shear flow is almost \(f_{D}=\sqrt{1/2}\).
A close look at the modes in the inset of
Fig.~\ref{fig:average_mixing_efficiency} reveals that the ptKD
characteristics were found.
For co-pump ptKDs, namely Fs-Ft and Bs-Bt, the modes of their PDFs
indicate that the history of the strain-rate type is planar shear flow.
In contrast, for counter-pump ptKDs, namely Fs-Bt and Bs-Ft, the
contribution of biaxial elongational flow at the pitched tips causes
mode to shift to a lower value of
\(\overline{\left(f_{D}\right)_{\alpha}}^{T_{\alpha}}\).
The result suggests that the counter-pump ptKDs to generate the
three-dimensional deformation, which promotes efficient mixing, are more
effectively than co-pump ptKDs.
\section{Concluding remarks}
For melt-mixing with a novel mixing element, the ``pitched-tip kneading
disk~(ptKD)'', its basic characteristics and mixing capability were
investigated based on numerical simulation of three-dimensional flow and the
statistics of the Lagrangian tracer.
The ptKD has two main geometric parameters: the pitched-tip
angle and the disk-stagger angle.
Four typical combinations of the pitched-tip angle and disk-stagger
angle were used in the current research.

The pressurization capability is mainly determined by the disk-stagger angle,
and the pitched-tip provides minor modification.

Flow patterns and transport mechanisms were studied according to the
tracer density evolution.  In co-pump ptKDs~(Fs-Ft and Bs-Bt), pipeline
flow exists that causes a large fluctuation in the Lagrangian history of
several quantities, including the exerted stress and the line-stretching
rate.
In co-pump ptKDs, the pitched-tip enhances the pressurization and causes
the axial transport to be inhomogeneous by generating pipeline flow.

In counter-pump ptKDs~(Fs-Bt and Bs-Ft), the biaxial elongational flow
occurs in front of the pitched-tip and enhances mixing over the channel.
The result suggests that the main advantage of Fs-Bt and Bs-Ft is gross
distributive mixing.
It is expected that, the optimal balance of distributive mixing and
dispersive mixing are found for various extrusion processes by adjusting
the combinations of the pitched-tip angle and the disk-stagger angle in
Fs-Bt and Bs-Ft.

To quantify of mixing capability, the maximum line-stretching rate and
the strain-rate-type identifier were introduced.
Information on the flow patterns and mixing capability is reflected in
the statistics of Lagrangian-history averages of the maximum
line-stretching rate and the strain-rate-type identifier.  These
quantities are useful for analyzing the results of three-dimensional
flow simulations.

Other characterizations of mixing in three-dimensional flow using
the Lyapunov exponent~\citep{lawal95:_mechan_of_mixin_in_singl} and
intensity of segregation~\citep{lawal95:_simul_of_inten_of_segreg} have
been suggested.
It would be interesting to compute these measures in ptKDs and to
compare the results with those of other kneading elements.
We are currently working on this project.

\section*{Acknowledgments}
The computation was partly carried out using the computer facilities at
the Research Institute for Information Technology in Kyushu University
and the Supercomputing Division of Information Technology Center at the
University of Tokyo.

\appendix
\section{Local line-stretching rate and type of strain-rate field}
\label{sec:def_flow_type_stretching_rate}
The mixing ability of a flow field can be quantified by the line-stretching
rate of an infinitesimally short line element \(\vec{l}\).
The time evolution of \(l\) obeys for a short time period
\begin{align}
\frac{\upd \vec{l}}{\upd t} &=\vec{l}\cdot\tensor{D},
\label{eq:dl_dt}
\end{align}
The stretching rate of the norm \(\left|\vec{l}\right|\) is derived 
from Eq.(\ref{eq:dl_dt}) as
\begin{align}
\frac{1}{\left|\vec{l}\right|}
\frac{\upd \sqrt{\left|\vec{l}\right|^{2}}}{\upd t} 
&=\frac{\vec{l}}{\left|\vec{l}\right|}\cdot
\frac{\upd \vec{l}}{\upd t}
\nonumber
\\
&=
\frac{\vec{l}}{\left|\vec{l}\right|}
\cdot\tensor{D}
\cdot
\frac{\vec{l}}{\left|\vec{l}\right|}
,
\label{eq:stretching_rate}
\end{align}
When the line-stretching rate
\(\tensor{D}:\vec{l}\vec{l}/\left|\vec{l}\right|^{2}\) is positive, an
infinitesimally small distance nearby two points separate in exponential
manner.

The stretching rate of \(\left|l\right|\) (\ref{eq:stretching_rate})
depends on the direction of \(\vec{l}\).  To characterize the mixing
ability of a given flow field, the maximum of
Eq.(\ref{eq:stretching_rate}) or equivalently the maximum
line-stretching rate, \(\lambda_{M}\) in
Eq.(\ref{eq:maximum_stretching_rate}) can be used. By definition,
\(\lambda_{M}\) coincides with the largest eigenvalue of \(\tensor{D}\).

In the theory of dynamical systems in a bounded system, the average
stretching rate based on the director advected by Eq.(\ref{eq:dl_dt}) on
the limiting trajectory is called the Lyapunov exponent, and a positive
Lyapunov exponent indicates that the dynamical system is
chaotic~\citep{ott02:_chaos_in_dynam_system}.
The Lagrangian-history average of
\(\overline{\left(\lambda_{M}\right)_{\alpha}}^{T_{\alpha}}\) is
different from the Lyapunov exponent in the choice of the director, but
can be used to characterize potential ability of line stretching by the
flow.
From the viewpoint of distributive mixing, the magnitude of
\(\lambda_{M}\) and its distribution in time and space can be used to
characterize the mixing ability.

While the line-stretching rate characterizes the stretching in one direction,
the strain rate tensor describes the tri-axial straining states.
By using the maximum line-stretching rate and \(\tensor{D}\), the types
of flow, namely the planar shear flow, uniaxial flow and biaxial
elongational flows, are identified.
The ratio of the maximum line-stretching rate to the second moment of
\(\tensor{D}\) is introduced in Eq.(\ref{eq:def_flow_type_indicator}).
It is noted that \(f_{D}\) coincides with the mixing
efficiency~\citep{ottino89:_kinem_of_mixin} except for the choice of the
direction \(\vec{n}\).
By definition,  \(f_{D}\) is an invariant of \(\tensor{D}\).
For incompressible flow, 
\(\tensor{D}:\tensor{D}\) is expressed by its largest and smallest
eigenvalues, \(\lambda_{M}\) and \(\lambda_{m}\),
\begin{align}
 \tensor{D}:\tensor{D}&=\lambda_{M}^{2}+\lambda_{m}^{2}+\left(\lambda_{M}+\lambda_{m}\right)^{2},
\end{align}
therefore \(f_{D}\) is found to be a function of \(\lambda_{m}/\lambda_{M}\)
as
\begin{align}
 f_{D}&=\frac{1}{\sqrt{
2\left[
1+\left(\frac{\lambda_{m}}{\lambda_{M}}\right)
+\left(\frac{\lambda_{m}}{\lambda_{M}}\right)^{2}
\right]
}},
\end{align}
The incompressibility condition indicates that \(\lambda_{M}>0\),
\(\lambda_{m}<0\), and the range of \(\lambda_{m}/\lambda_{M}\in [-2,-1/2]\).
Thus \(f_{D}\) is increasing monotonically on
\(\lambda_{m}/\lambda_{M}\) and lies in
\begin{align}
 \sqrt{\frac{1}{6}}\leq f_{D} \leq \sqrt{\frac{2}{3}}
 &~~\text{for}~~(-2\leq \frac{\lambda_{m}}{\lambda_{M}}<-\frac{1}{2}),
\end{align}
For uniaxial and biaxial elongational flows,
\(\lambda_{m}/\lambda_{M}=-1/2\) and \(\lambda_{m}/\lambda_{M}=-2\),
respectively.  For planar shear flow, \(\lambda_{m}/\lambda_{M}=-1\).
The corresponding \(f_{D}\) is summarized in
Eq.(\ref{eq:flow_type_indicator}).

Because the uniaxial elongational flow establishes a more effective flow
pattern in dispersive mixing than simple shear flow, several
authors~\citep{yang92:_flow_field_analy_of_knead,yao98:_influen_of_desig_disper_mixin,cheng97:_study_of_mixin_effic_in,cheng98:_distr_mixin_in_convey_elemen}
have utilized some indicator of the strain pattern based on
\(\tensor{D}\) and \(\tensor{\Omega}\) like
\begin{align}
\lambda &= \frac{
\sqrt{\tensor{D}:\tensor{D}}
}{
\sqrt{\tensor{D}:\tensor{D}}
+
\sqrt{\tensor{\Omega}^{T}:\tensor{\Omega}}
},
\\
S_{f} &= \frac{
2\left(\tensor{D}:\tensor{D}\right)^{2}
}{
\stackrel{\circ}{\tensor{D}}:
\stackrel{\circ}{\tensor{D}}
},
\end{align}
where \(\tensor{\Omega}\) is the vorticity tensor and
\(\stackrel{\circ}{(.)}\) indicates the steady co-rotational time
derivative.
Both \(\lambda\) and \(S_{f}\) takes certain values for elongational
flow, simple shear flow, and simple rotational flow. The flows can be
discriminated based on the ratio of the a magnitude of vorticity
\(\tensor{\Omega}\) to a magnitude of \(\vec{\nabla}\vec{v}\) which
indirectly identifies the strain pattern:
null vorticity, \(\tensor{\Omega}=\tensor{0}\),
in pure elongational flow,
and \( \tensor{D}:\tensor{D}=\tensor{\Omega}^{T}:\tensor{\Omega}\) in
simple shear flow. 
By definition, \(\lambda\) and \(S_{f}\) cannot
distinguish
between uniaxial and biaxial elongational flows.
Like \(\lambda\) and \(S_{f}\), \(f_{D}\) can be used to identify a type
of strain-rate. 
It is noted that \(f_{D}\) is different from \(\lambda\) and \(S_{f}\)
in that \(f_{D}\) is based solely on the strain rate \(\tensor{D}\) and
provides a direct characterization of \(\tensor{D}\) because we are
interested in the stretching rate of the distance of two nearby points
and the rotational component of flow does not affect the maximum
line-stretching rate.

\end{document}